\documentclass{article}

\usepackage{arxiv}

\usepackage[utf8]{inputenc} 
\usepackage[T1]{fontenc}    
\usepackage{hyperref}       
\usepackage{url}            
\usepackage{booktabs}       
\usepackage{amsfonts}       
\usepackage{nicefrac}       
\usepackage{microtype}      
\usepackage{lipsum}		
\usepackage{graphicx}
\usepackage{doi}
\usepackage{threeparttablex}
\usepackage{sistyle}
\SIthousandsep{,}

\usepackage{import}

\def\bbeta{{\boldsymbol{\beta}}}

\usepackage{amsthm}

\usepackage{booktabs}
\usepackage{amsmath}

\newcommand{\betahat}{\hat{\beta}}
\newcommand{\bbetahat}{\hat{\pmb{\beta}}}

\usepackage{mathtools}
 
\usepackage{dsfont}

\title{Overcoming data challenges through enriched validation and targeted sampling to measure whole-person health in electronic health records}

\author{ \href{https://orcid.org/0000-0001-5380-2427}{\includegraphics[scale=0.06]{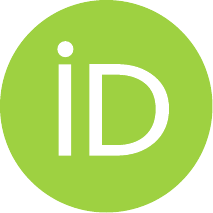}\hspace{1mm}Sarah C.~Lotspeich} \\
	Department of Statistical Sciences\\
	Wake Forest University\\
	Winston-Salem, NC 27109 \\
	\texttt{lotspes@wfu.edu} \\
	\And
	Sheetal ~Kedar \\
	Department of Anesthesiology\\
    Wake Forest University School of Medicine \\
	Winston-Salem, NC 27157 \\
    \And
    Rabeya ~Tahir \\
	Department of Anesthesiology\\
    Wake Forest University School of Medicine \\
	Winston-Salem, NC 27157 \\
    \And
    Aidan D. ~Keleghan \\
	Department of Anesthesiology\\
    Wake Forest University School of Medicine \\
	Winston-Salem, NC 27157 \\
    \And
    Amelia ~Miranda \\
	Department of Anesthesiology\\
    Wake Forest University School of Medicine \\
	Winston-Salem, NC 27157 \\
    \And
    Stephany N. ~Duda \\
	Department of Biomedical Informatics\\
	Vanderbilt University Medical Center \\
	Nashville, TN 37203 \\
    \And 
    Michael P. ~Bancks \\
	Department of Epidemiology and Prevention\\
	Wake Forest University School of Medicine \\
	Winston-Salem, NC 27157 \\
    \And 
    \href{https://orcid.org/0000-0001-7310-6525}{\includegraphics[scale=0.06]{orcid.pdf}\hspace{1mm}Brian J. ~Wells} \\  
    Department of Biostatistics and Data Science\\
	Wake Forest University School of Medicine \\
	Winston-Salem, NC 27157 \\
    \And 
    \href{https://orcid.org/0000-0002-9083-891X}{\includegraphics[scale=0.06]{orcid.pdf}\hspace{1mm}Ashish K. ~Khanna} \\ 
    Department of Anesthesiology,\\ Division of Critical Care Medicine, \\
    Wake Forest University School of Medicine \\
	Winston-Salem, NC 27157 \\
    Outcomes Research Consortium\\
	Houston, TX 77030 \\
    \And 
    \href{https://orcid.org/0000-0001-6265-0752}{\includegraphics[scale=0.06]{orcid.pdf}\hspace{1mm}Joseph ~Rigdon} \\ 
    Department of Biostatistics and Data Science\\
	Wake Forest University School of Medicine \\
	Winston-Salem, NC 27157 \\
}

\renewcommand{\shorttitle}{Overcoming data challenges to measure whole-person health in electronic health records}

\hypersetup{
pdftitle={ALI in EHR},
pdfsubject={q-bio.NC, q-bio.QM},
pdfauthor={Sarah C.~Lotspeich, Sheetal ~Kedar, Rabeya ~Tahir, Aidan D. ~Keleghan, Amelia ~Miranda, Stephany N. ~Duda, Michael P. ~Bancks, Brian J. ~Wells, Ashish K. ~Khanna, Joseph ~Rigdon},
pdfkeywords={chart review, data audits, healthcare utilization, measurement error, missing data, outcome-dependent sampling, two-phase designs},
}

\begin{document}
\maketitle

\begin{abstract}
\textit{Objective:} The allostatic load index (ALI) is a 10-component composite measure of whole-person health, which reflects the multiple interrelated physiological regulatory systems that underlie healthy functioning. Data from electronic health records (EHR) present a huge opportunity to operationalize the ALI in learning health systems; however, these data are prone to missingness and errors. Validation (e.g., through chart reviews) can provide better-quality data, but realistically, only a subset of patients’ data can be validated, and most protocols do not recover missing data. \\ 
\textit{Methods:} Using a representative sample of $1000$ patients from the EHR at an extensive learning health system ($100$ of whom could be validated), we propose methods to design, conduct, and analyze statistically efficient and robust studies of ALI and healthcare utilization. Employing semiparametric maximum likelihood estimation, we robustly incorporate all available patient information into statistical models. Using targeted design strategies, we examine ways to select the most informative patients for validation. Incorporating clinical expertise, we devise a novel validation protocol to promote EHR data quality and completeness.  \\
\textit{Results:} Chart reviews uncovered few errors ($99\%$ matched source documents) and recovered some missing data through auxiliary information in patients' charts. On average, validation increased the number of non-missing ALI components per patient from $6$ to $7$. Through simulations based on preliminary data, residual sampling was identified as the most informative strategy for completing our validation study. Incorporating validation data, statistical models indicated that worse whole-person health (higher ALI) was associated with higher odds of engaging in the healthcare system, adjusting for age. \\
\textit{Conclusion:} Targeted validation with an enriched protocol can ensure the quality and promote the completeness of EHR data. Findings from our validation study were incorporated into analyses as we operationalize the ALI as a scalable whole-person health measure that predicts healthcare utilization in the learning health system. 
\end{abstract}


\section{Introduction}
\label{sec:motivation} 

\subsection{Measuring Whole-Person Health} 

The advent of learning health systems \cite{Easterling2022,Greene2012} signals a transition from a reactive (treatment-based) to a proactive (prevention-based) philosophy in healthcare. This transition could prevent future healthcare utilization through early intervention, improving individual health and providing cost savings to the institution. Having a standardized summary measure of health that is simple to compute and track could play a central role in disease prevention (saving lives) and reducing healthcare utilization (saving costs).

Having easy access to a standardized summary measure of whole-person health in electronic health records (EHR) would be advantageous for patients, clinicians, and researchers \cite{Khurana2022, WholePersonHealth}. The electronic frailty index (EFI) serves this purpose but only for those aged $65$ and older \cite{Pajewski2019}. A whole-person health metric for adults aged $18$ and older would enable system-wide longitudinal tracking of individual and population health, comparisons across geographies and hospitals, and integration of an important predictor in algorithms relevant to the enterprise (e.g., predicting readmissions). 

Individual physiological variables only measure aspects of health. The allostatic load index (ALI) is a composite measure of whole-person health, which reflects the multiple interrelated physiological regulatory systems that underlie healthy functioning \cite{Beckie2012, McEwen1999, Nobel2017, Seeman2001}. It has been used to predict numerous health outcomes (e.g., \cite{Hux2012, Juster2011, Seplaki2004, Shalowitz2019}) and measure racial health disparities \cite{Geronimus2006}. Because it is sensitive to short-term interventions, ALI is also an appealing ``surrogate'' outcome to test health interventions.

Many ALI formulations exist. We focus on the Seeman et al. (2001) \cite{Seeman2001} calculation, which computes ALI from ten components across three body systems (Table~\ref{tab:ali_components_roadmap}). The original numeric measurements (e.g., hemoglobin A1c [HbA1c]) are discretized at clinically-driven thresholds to define binary indicators of being at an unhealthy level (e.g., $= 1$ if HbA1c $\geq 6.5$ and $= 0$ otherwise). ALI is the sum of these ten components, with zero and ten representing the best and worst whole-person health, respectively (Supplemental Figure~S1). 

Whole-person health is a latent concept influenced by genetics, environment, and other factors. Scores like the ALI yield ``computable phenotypes'' for it, concisely characterizing patients' current health status with one-number summaries of observed data \cite{Halpern2014, Ting2023}. The ALI's ability to quantify whole-person health has been rigorously examined in controlled studies \cite{Beckie2012, McEwen1999, Seeman2001}. However, before adopting the ALI in other settings, it must be revalidated accounting for real-world data challenges. 

\subsection{Data Quality Challenges in Electronic Health Records Data}

EHR data present a substantial opportunity to operationalize the ALI in the learning health system. The volume of routinely collected EHR data is steadily climbing, increasing the accessibility to clinically meaningful variables. Analyzing data that come at little cost for collection is particularly appealing to researchers, leading to a surge in EHR data use in many clinical domains (e.g., \cite{Green2013, Tannen2009, Wei2015, zaniewksi2018}).  

However, EHR data are collected in the fast-paced world of clinical care, leading to missingness and errors. Quality concerns remain a significant hurdle to the large-scale adoption of EHR in healthcare research \cite{Hersh2013, Kim2019, Nordo2019, Wells2014}. In calculating the ALI, measurements may be (i) missing, making the component missing, or (ii) non-missing but mismeasured, misclassifying the component. 

\begin{figure}[ht]
\centering
\includegraphics[width=\textwidth]{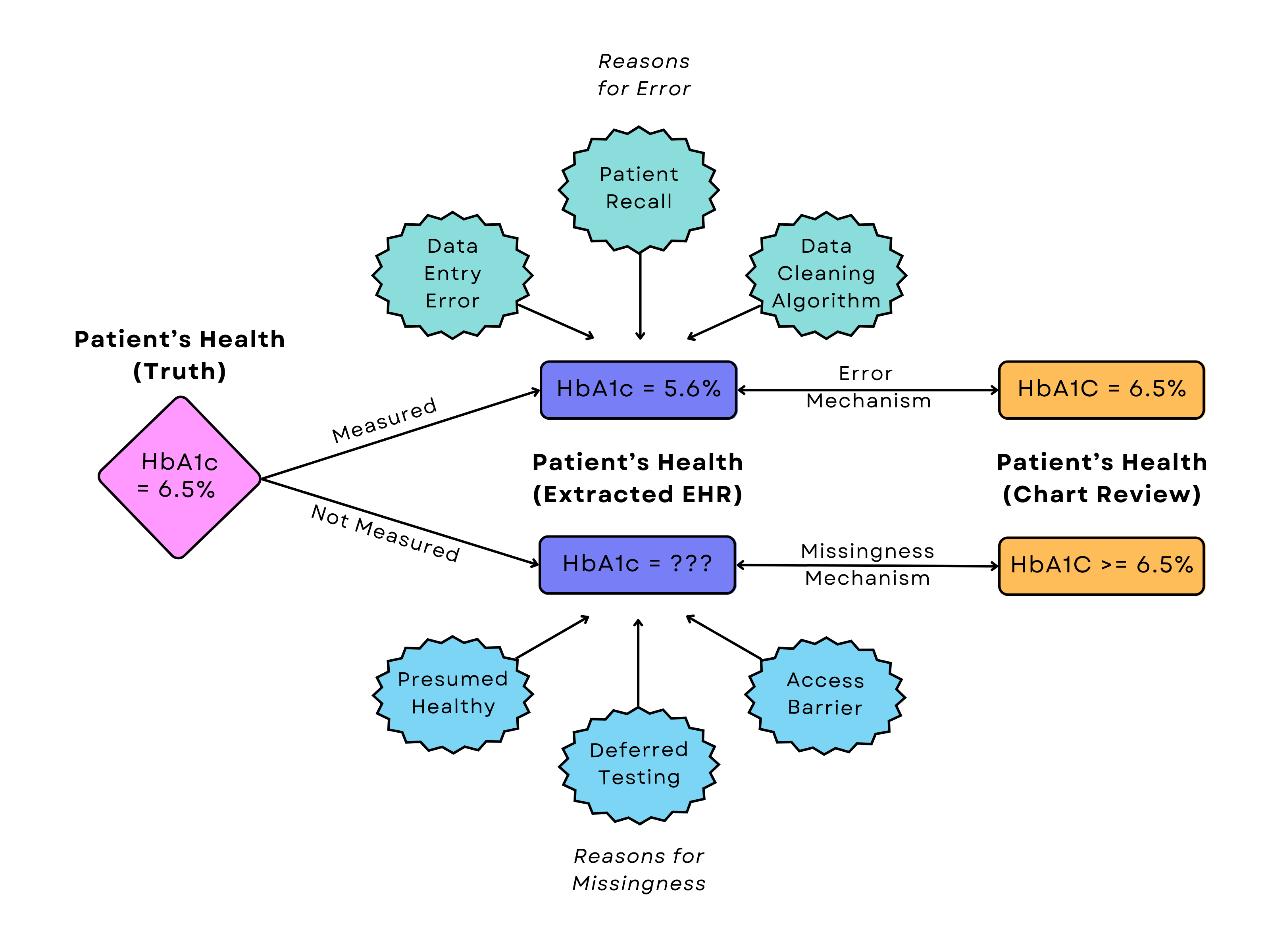}
\caption{Diagram of potential errors and missingness in a patient's hemoglobin A1c (HbA1c) measurement in the electronic health records (EHR) before and after validation (chart review). In the top scenario, the patient had HbA1c measured but with error, and the correct value was recovered through validation. In the bottom scenario, the patient did not have HbA1c measured, but auxiliary information was located through validation to replace the missing component.}\label{ALI-flow}
\end{figure}

Consider a patient whose true HbA1c was $6.5\%$, so their corresponding component should equal one. We illustrate two scenarios for this patient's data in Figure~\ref{ALI-flow}. First, suppose HbA1c was measured but erroneously transposed to $5.6\%$; their component would be misclassified as zero, since $5.6\%$ falls into the healthy range. Second, suppose HbA1c was not measured, so the value is missing. Then, we cannot define their component. Counting missing components as zeros would also induce misclassification by treating them as ``healthy." We accommodate missing data by instead defining ALI as the \textit{proportion} of non-missing components a patient experienced. 

Ignoring errors and missingness can hinder analyses, biasing inference \cite{ Chen2019, Duan2017} and reducing predictive accuracy \cite{Wang2016}. There is particular concern for derived variables, like ALI, that combine multiple variables \cite{Beesley2022,Sinnott2014}. Analyzing the ALI computed from EHR data ``as is'' may misrepresent a patient’s actual health and lead to false conclusions about its association with healthcare utilization. 

Applying popular missing data approaches like multiple imputation (MI) may offer some improvement at baseline \cite{Wells2013}. However, they can also hinder results if any imputation models are misspecified \cite{Noghrehchi2021}. Since components are likely missing not at random in EHR data (e.g., clinicians usually order labs when they suspect abnormal values) \cite{Haneuse2021}, correctly specifying imputation models could be especially challenging. Plus, MI would not correct for  errors. 

For the ALI to be useful, it must be reliably and accurately measured. The original calculations, like Seeman et al. (2001) \cite{Seeman2001}, were done in clinical trials, where all patients \textit{must} have all measurements. Further, data validation protocols were likely in place to ensure quality. Similarly, validation via chart review, comparing extracted EHR data to source documents (like electronic charts), provides data quality insights. Complete data validation (for all patients and variables) would ensure the integrity of EHR data but is expensive and unattainable. 

Fortunately, partially validating EHR data can identify issues and pave the way for corrections. It provides a cost-effective alternative to complete data validation and is a promising approach to gauge the quality of observational data sources \cite{Lotspeich2020, Lotspeich2023, shepherd2022}. Partial validation studies are two-phase designs \cite{Tao2019, White1982}. Phase I includes EHR data for $N$ patients, which can be used to select a subsample of $n$ patients ($n < N$) for validation (chart reviews) in Phase II.

Validated data are usually expected to be more correct (i.e., less error-prone) but not more complete (i.e., less missingness). To calculate the ALI from EHR data, we needed validation to do both. We devise a novel validation protocol that incorporates a ``roadmap'' created by clinical experts to recover missing data using auxiliary  information (e.g., related diagnoses). With it, we verify the ALI's strength as a computable phenotype for whole-person health and a predictor of healthcare utilization in observational settings with data challenges.

\begin{table}[ht!]
\centering
\resizebox{\columnwidth}{!}{
\begin{tabular}{rlcc}
        \textbf{System} & \textbf{Component}& \textbf{Threshold} & \textbf{Auxiliary Information} \\
        \hline 
        Cardiovascular & Systolic Blood Pressure & $>140$ & Hypertension  \\
        & Diastolic Blood Pressure & $>90$ & Hypertension  \\
        Metabolic & Body Mass Index & $>30$ & Obesity; Morbid Obesity; \\
        & & & Grade I, II or III Obesity\\
        & Triglycerides & $\geq 150$ &	Hypertriglyceridemia \\
        & Total Cholesterol & $\geq 200$ & Hypercholesterolemia \\
        Inflammation & C-Reactive Protein & $\geq 10$	& Sepsis; Infection; Auto-Immune \\
        & & & Inflammatory Syndrome\\
        & Hemoglobin A1C & $\geq 6.5$	& Diabetes; Impaired \\
        & & & Glycemic Control \\
        & Serum Albumin & $\geq 3.5$ & \textit{(None Given)} \\
        & Creatinine Clearance & $<110$ (Males)& Renal Failure; \\
        & & $<100$ (Females)  & Insufficiency; Acute \\
        & & & Kidney Injury; \\
        &&& Chronic Renal Failure
        \\
        & Homocysteine & $>50$ &	Hyperhomocysteinemia; \\
        & & & Vitamin deficiency \\
    \end{tabular}
}
\caption{Ten components of the allostatic load index (ALI) were defined by discretizing measurements across three body systems at clinically-driven thresholds. If the component was missing from the extracted electronic health records (EHR) data, auditors searched for auxiliary information in the patient's medical chart in Epic (the institution's electronic charting software). If the auxiliary information was present, the patient was treated as having an ``unhealthy'' measurement and the component was included in their validated ALI.}
\label{tab:ali_components_roadmap}
\end{table}

\subsection{Application to the Learning Health System}

We investigate how to overcome data challenges to measure whole-person health across one of the largest learning health systems in the United States. Having recently integrated the EFI into the EHR to measure whole-person health for patients aged $65$ and older \cite{Orkaby2024}, the health system is interested in operationalizing the ALI for the broader patient population. We obtained a representative sample of $N=1000$ patients from the EHR at Atrium Health Wake Forest Baptist Hospital in Winston-Salem, North Carolina. Our sample was drawn from patients (i) $18$--$65$ years old who (ii) initiated their first primary care outpatient encounter in the health system between March 11, 2018 and March 10, 2020. Outpatient encounters were chosen to represent patients voluntarily engaging in care. 

Our data contained demographics, vitals, and lab values dating back $3$ years before the first encounter. Key variables included measurements for the ten ALI components (Table~\ref{tab:ali_components_roadmap}), age at first encounter (hereafter, ``age''), and an indicator of having at least one hospitalization or emergency department (ED) visit during the $2$-year study period. Since most patients had multiple encounters, ALI components were based on mean measurements. 

\subsection{Overview}

While very insightful in the development and use of computable phenotypes, partial validation studies are expensive, time-intensive undertakings. Using a representative sample of $1000$ patients from the EHR at an extensive academic learning health system ($100$ of whom could be validated), we propose novel methods to design, conduct, and analyze statistically efficient and robust studies of whole-person health (ALI) and healthcare utilization. Using statistics and biomedical informatics methods, we seize three key opportunities to maximize our investment into partial validation: robust and efficient statistical modeling, enriched chart review protocol, and targeted sampling design. \\

\noindent \textbf{Statement of Significance}

\begin{itemize}
    \item[]\textbf{Problem} Standardized summary measures of whole-person health are needed in  the EHR to make patient- and institution-level decisions. However, routinely collected data are prone to missingness and errors. 
    \item[] \textbf{What Is Already Known} Validation, wherein extracted EHR data are compared to clinical source documents (like electronic charts), provides insights into data quality, but validating all patients is not feasible, and existing protocols only correct for errors, not missingness. 
    \item[] \textbf{What This Paper Adds} We devise a robust and efficient model for partially validated, error-prone EHR data; propose a targeted validation study design to identify the most informative patients for our model; and develop an enriched validation protocol that incorporates clinical expertise to improve the quality \textit{and} completeness of EHR data. 
\end{itemize}

\section{Methods}\label{sec:meth}

\subsection{Healthcare Utilization Model} \label{sec:meth_model} 

This study was approved by the Institutional Review Board at Wake Forest University School of Medicine (WFUSM). We model healthcare utilization ($Y\in\{0, 1\}$), defined as a binary indicator of having at least one hospitalization or ED visit in the $2$-year study period. Whole-person health, measured by ALI ($X\in[0, 1]$), is the exposure of interest. Each component $S_j$ ($j\in\{1, \dots, 10\}$) equals one if the patient's measurement was considered unhealthy and zero otherwise, such that $X=\sum_{j=1}^{10}S_{j}/10$ is the proportion of unhealthy measurements experienced. Further, we control for age ($Z \in \mathcal{R}$). 

We assume a logistic regression model for the log-odds that $Y=1$ given ($X$, $Z$). A sample of $N$ patients was extracted from the EHR to estimate these log odds ratios (logORs), denoted by $\bbeta$. Age $Z$ is fully-observed and expected to be error-free. However, only an error-prone measure for $X$, denoted by $X^*$ ($X^*\in[0, 1]$), is available from the EHR data. We define error-prone components $S_j^*$ ($j\in\{1, \dots, 10\}$)  and introduce missingness indicators  $M_j^*$ for them (i.e., $M_j^*=1$ if $S_j^*$ is missing). We compute $X^* = \sum_{j=1}^{10}(1 - M_j^*)S_{j}^*/\sum_{j=1}^{10}(1 - M_j^*)$, i.e., as the proportion of \textit{non-missing} components experienced. Calculating ALI as a proportion was critical. The sum of non-missing components treats missing values as zeros, whereas the proportion ignores them. 

\subsection{Three Key Steps to Partial Validation Studies: Analysis, Design, and Protocol}\label{subsec:key_steps} 

To consistently estimate the healthcare utilization model, we planned a partial validation study. We extracted EHR data ($Y$, $X^*$, $Z$) for a sample of $N=1000$ patients. A subset of $n=100$ patients was selected for validation (chart review) such that $X$ was also measured.

\begin{figure}[ht]
\includegraphics[width=\textwidth]{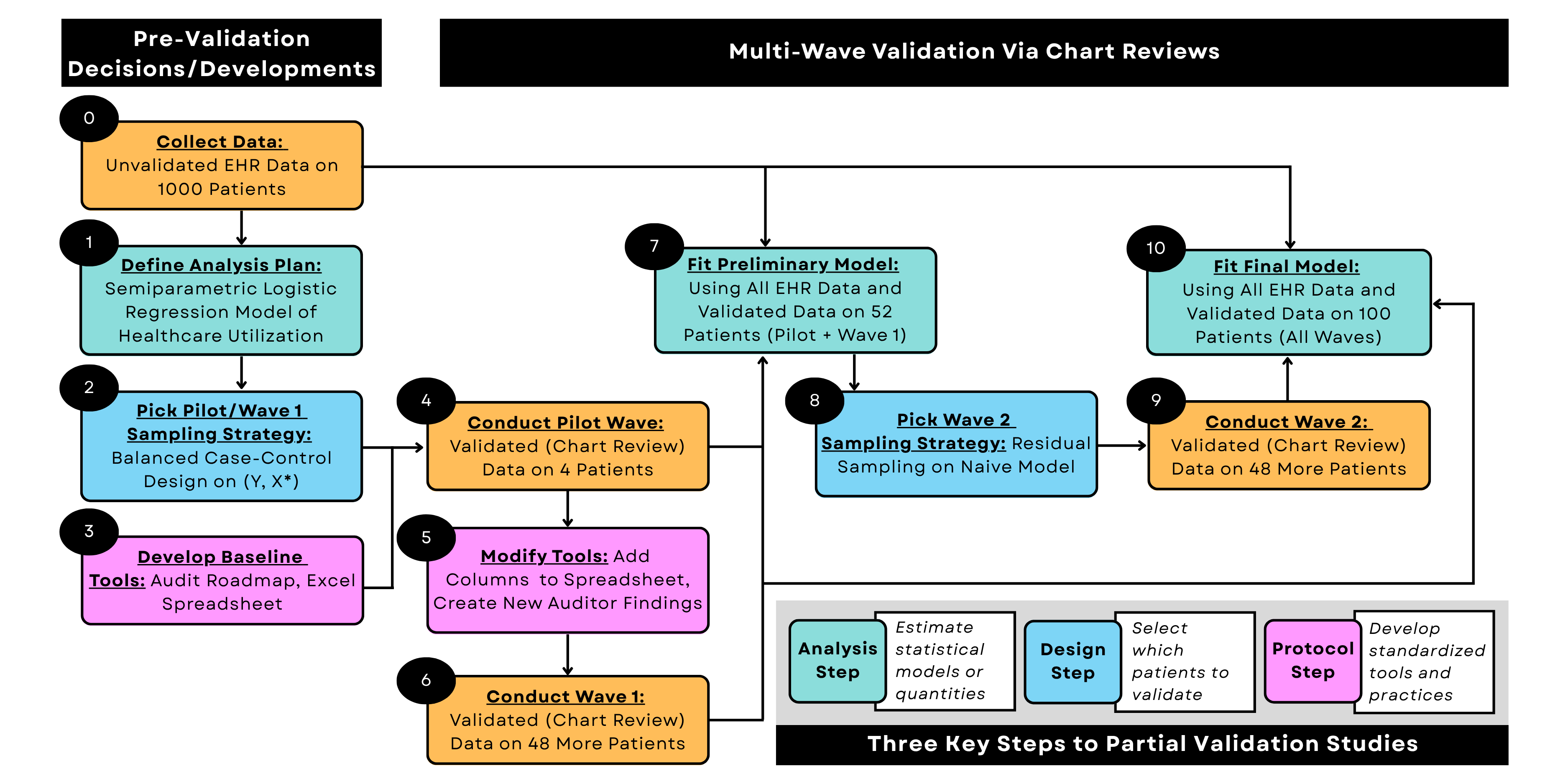}
\caption{After extracting the electronic health records (EHR) data, we prepared for validation by defining our analysis plan, picking an initial sampling strategy, and developing baseline validation tools. We first conducted the Pilot Wave chart reviews. This information was then used to modify the validation protocol and establish inter-auditor agreement. Next, the Wave I chart reviews followed the same sampling strategy as the Pilot but used the updated validation tools. With the EHR data ($N = 1000$) and chart review data from the Pilot and Wave I ($n = 52$), we fit a preliminary version of the healthcare utilization model. Estimates from this model guided our selection of the targeted sampling strategy for Wave II via bespoke simulation studies. Following completion of all validation waves, we combined the EHR data ($N = 1000$) with all the chart review data ($n = 100$) to fit the final healthcare utilization model.}\label{partial-validation-steps}
\end{figure}

Partial validation studies break down into three key steps. \textit{Design:} We select which patients should be validated. \textit{Protocol:} We develop standardized tools and practices for those patients' chart reviews. \textit{Analysis:} We estimate statistical models from validated and unvalidated data together. These steps are intertwined, with design choices depending on the planned analysis, and each step provides opportunities to gain information. For added flexibility, a multi-wave validation framework (i.e., conducting chart reviews in multiple sprints) allows us to revisit key steps and make modifications as information accumulates (Figure~\ref{partial-validation-steps}).

We built the statistical model and developed the baseline sampling strategy and tools before beginning validation. The proposed methods are detailed in order of development: analysis, design, and protocol. Across three validation waves, we updated our sampling strategy and tools (revisiting the design and protocol steps, respectively) before fitting the final model. 

\subsection{{Analysis: Robust and Efficient Statistical Model for Partially Validated Data}} \label{sec:meth_smle} 

Let $V_i$ distinguish validated and unvalidated patients, where $V_i=1$ if patient $i$ is validated ($i\in\{1, \dots, N\}$) and $V_i=0$ otherwise. Incorporating EHR and chart review data, along with the validation indicator, the joint distribution of a complete observation is assumed to be 
\begin{align}
\Pr(V,X^*,Y,X,Z) &= \Pr(V\mid  Y,X^*,Z){\Pr}_{\bbeta}(Y\mid X,Z)\Pr(X\mid  X^*,Z)\Pr(X^*,  Z), \label{joint}
\end{align}
where $\Pr(V\mid  Y,X^*,Z)$ is the validation sampling probability; ${\Pr}_{\bbeta}(Y\mid  X,Z)$ is the logistic regression model; $\Pr(X \mid  X^*,Z)$ is the exposure error mechanism; and $\Pr(X^*,  Z)$ is the joint probability density or mass function of ($X^*$, $Z$). Two assumptions are made in \eqref{joint}. First, validation sampling depends only on fully-observed variables ($Y$, $X^*$, $Z$), so $X$ is missing at random (MAR) for unvalidated patients \cite{Little1992}. Second, $X^*$ is a surrogate for $X$, such that $Y$ and $X^*$ are conditionally independent given $X$.

Observations for all $N=1000$ patients are assumed to be independent and identically distributed following \eqref{joint}. Thus, the observed-data log-likelihood function (hereafter, log-likelihood) for the logORs $\bbeta$ is defined from \eqref{joint} after two simplifications. First, we can omit $\Pr(V|Y,X^*,Z)$ from the log-likelihood because $X$ is MAR. Second, we can ignore $\Pr(X^*, Z)$ because it drops out as the log-likelihood is maximized with respect to $\bbeta$, since ($X^*$, $Z$) are fully-observed. Now, the log-likelihood for $\bbeta$ is proportional to
\begin{align}
&\sum_{i=1}^{N}V_i\left[\log\left\{{\Pr}_{\bbeta}(Y_i\mid X_i,Z_i)\right\}+\log\left\{\Pr(X_i\mid X_i^*,Z_i)\right\}\right] \nonumber \\
& + \sum_{i=1}^{N}(1-V_i)\log\left\{\sum_{x=0}^{1}{\Pr}_{\bbeta}(Y_i\mid x,Z_i)\Pr(x\mid X_i^*,Z_i)\right\}. \label{od_ll2}
\end{align}
Notice that unvalidated patients  contribute  incomplete information  to \eqref{od_ll2} by summing over the missing validated $X$: 
\begin{align}
\sum_{x=0}^{1}{\Pr}_{\bbeta}(Y_i\mid x,Z_i)\Pr(x\mid X_i^*,Z_i) &\propto \sum_{x=0}^{1}{\Pr}_{\bbeta}(Y_i, x,Z_i,X_i^*) = {\Pr}_{\bbeta}(Y_i,Z_i,X_i^*), \label{int_joint}
\end{align}
where $X$ takes on values $x \in \{0, 0.1, \dots, 0.9, 1\}$ here. 

A model for $\Pr(X\mid X^*,Z)$ is needed to estimate $\bbeta$ from \eqref{od_ll2}, since $X$ is only observed for validated patients. Since $X\in[0,1]$, there are not many flexible, parametric choices to model this conditional distribution. A nonparametric estimator is preferable, allowing $X$ to take on any values and arbitrarily depend on ($X^*$, $Z$). 

Lotspeich et al. (2022) \cite{lotspeichetal2022} proposed semiparametric sieve maximum likelihood estimators (SMLEs) for logistic regression with outcome and exposure errors. They nonparametrically handle the exposure error mechanism as described. A brief outline of how to adapt the SMLEs for exposure error only is provided in the Supplemental Materials. The \textit{logiSieve} package contains R functions to fit this model \cite{logiSieve}.

\subsection{Design: Targeted Sampling to Select Informative Patients for Validation}\label{sec:meth_design} 

\begin{figure}
    \centering \includegraphics[width=0.75\textwidth]{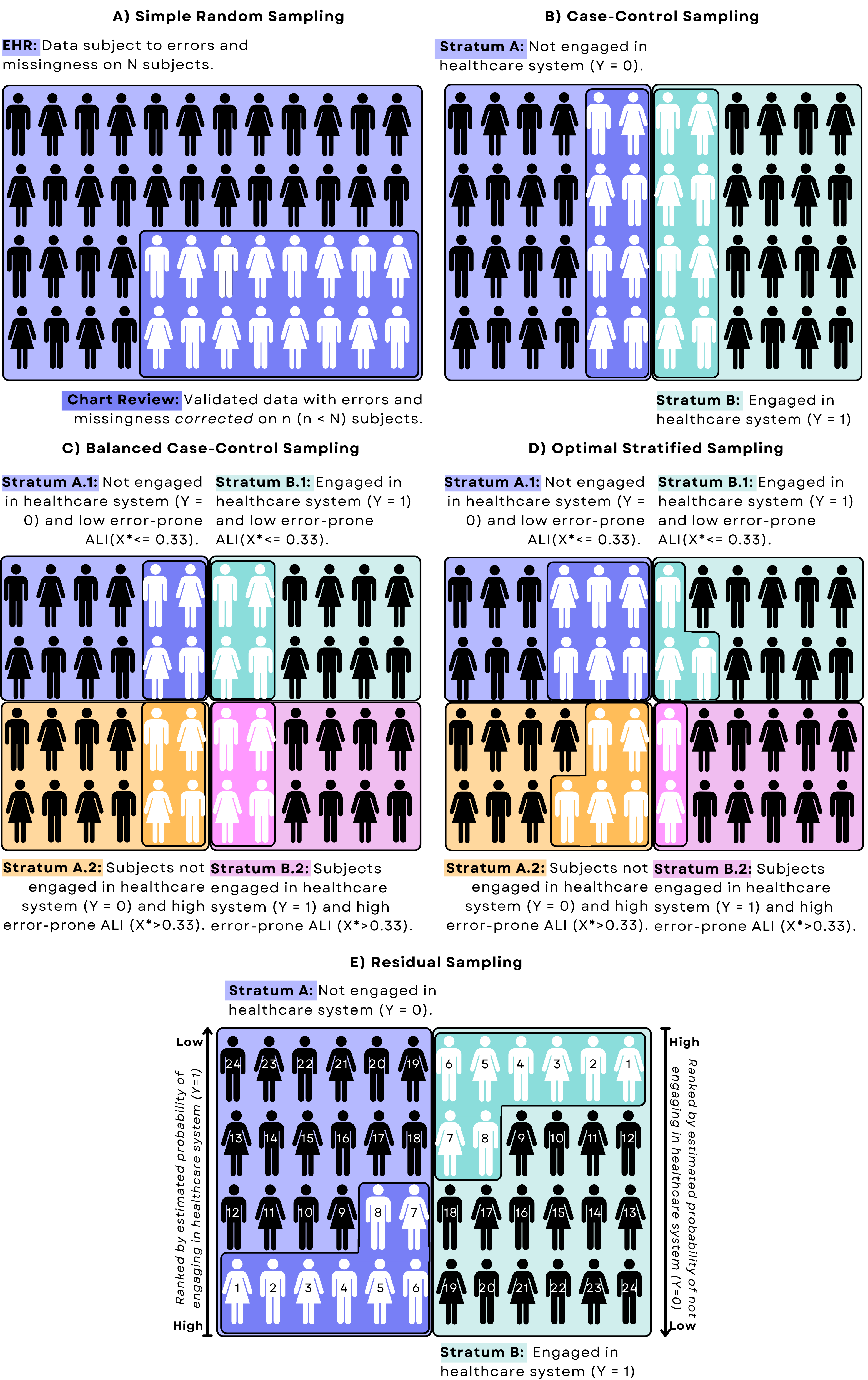}
    \caption{Illustrative diagram of partial validation studies with $N=48$ patients extracted from the electronic health records (EHR) and $n=16$ patients chosen for validation via chart review. The least informative strategy to select the validation study is \textbf{A)} simple random sampling. More informative targeted strategies include \textbf{B)} case-control sampling, \textbf{C)} balanced case-control sampling using discretized ALI, \textbf{D)} optimal stratified sampling using discretized ALI, and \textbf{E)} residual sampling based on the naive healthcare utilization model. Within strata, the darker shaded areas indicate which patients would be selected.}
    \label{diag-all-designs}
\end{figure}

Based on our analysis plan, we considered various sampling designs to select the most informative $100$ patients for validation. When validating computable phenotypes, \textit{simple random sampling (SRS)} is often used. Under SRS, $100$ patients would be chosen for chart review with equal probability (Figure~\ref{diag-all-designs}A). This strategy ignores all available information. Some phenotyping algorithms require an SRS validation study. However, for full-likelihood approaches like SMLEs, any information available for all patients can be used to design the validation study. Designing based on fully-observed information ensures that the MAR assumption holds for the downstream analysis while allowing for many targeted strategies. 

When designing a validation study to promote statistical efficiency of a parameter estimate (e.g., to minimize the variance of the estimated logOR $\hat{\beta}_1$  on ALI), we start at the end. Beginning with the planned logistic regression model of $Y$ given ($X$, $Z$), we work backward to determine which patients will be most informative. For this reason, the ``best'' designs vary across different estimators, with model-based ones (like the SMLEs) offering the most flexibility. Estimators that rely on the design to handle missingness (like inverse probability weighting [IPW]) require a nonzero sampling probability for all patients; model-based estimators do not. 

We considered stratified and rank-based designs. Multiple validation waves were conducted to allow the protocol and study design to adapt as information accumulated, which is becoming common practice in this space \cite{lotspeichetal2024, McIsaac&Cook2015, shepherd2022}. Ultimately, simulations based on preliminary validation data (Section~\ref{subsec:multiwave}) guided our selection of the targeted design used in the application. 

\subsubsection{Stratified Designs}

Given our binary outcome, the \textit{case-control (CC)} design is a natural option, wherein $50$ patients each are chosen with $Y=0$ and $Y=1$ (Figure~\ref{diag-all-designs}B). The \textit{balanced case-control (BCC)} design further stratifies on covariates and chooses equal (balanced) numbers from each stratum (Figure~\ref{diag-all-designs}C). Continuous covariates must be discretized to form strata. For instance, we could let $X_D^*$ define strata of patients with error-prone ALI $X^*$ above and below the average and then validate $25$ patients from each of the four ($Y$, $X_D^*$) strata. Importantly, we can still fit the model using the continuous values $X^*$. 

The CC and BCC designs are intuitive and offer efficiency gains over SRS. Still, slightly more complicated designs are guaranteed to be theoretically ``optimal;'' they choose the number of patients per stratum to achieve the lowest possible variance for $\hat{\beta}_1$ \cite{Breslow&Cain1988, Chen&Lumley2020, holcroft&spiegelman1999, McIsaac&Cook2014, Reilly&Pepe1995, Tao2019}. Lotspeich et al. (2024) \cite{lotspeichetal2024} proposed the \textit{optimal (OPT)} design for model-based analyses with binary outcome and exposure misclassification (Figure~\ref{diag-all-designs}D), which could apply here. However, we would achieve the lowest possible variance for $X_D^*$ rather than $X^*$. Increasing the number of unique values of $X_D^*$ (i.e., creating more strata) can help, but information will still be lost. Moreover, the OPT design depends on the true logORs $\bbeta$, which are unknown. Reliable placeholders for $\bbeta$ (often collected in a multi-wave framework) are required to implement the design.

\subsubsection{Rank-Based Designs}

Designs ranking patients offer improved granularity and efficiency gains. For example, \textit{extreme tail sampling (ETS)} designs order patients by some value of interest and target those with the smallest or largest (most ``extreme'') values. To identify patients who are most informative to the healthcare utilization model, \textit{residual sampling (RS)} is preferred \cite{Lin2013}. We begin by fitting the ``naive'' analysis model, i.e., using standard logistic regression procedures with $Y$, $X^*$ (instead of $X$), and $Z$ to estimate the logORs $\bbetahat^*$ from the full sample. Then, we compute residuals for all patients as 
\begin{align}
    r_i^* &= Y_i - {\Pr}_{\bbetahat^*}(Y_i=1|X^*_i,Z_i) \nonumber \\
    &= Y_i - \left[1 + \exp\left\{-\left(\betahat_0^* + \betahat_1^*X_i^* + \betahat_2^{*}Z_i\right)\right\}\right]^{-1}, \label{def_resid}
\end{align}
and sample the $100$ patients with the most extreme residuals (Figure~\ref{diag-all-designs}E). The largest residuals (in magnitude) correspond to patients whose observed $Y_i$ had a low predicted probability. The \textit{auditDesignR} package contains R functions to sample from all of these designs \cite{auditDesignR}. 

\subsection{{Protocol: Enriched Validation Protocol to Reduce Errors and Missingness}} \label{sec:meth_protocol}

Chart reviews were conducted over $6$ months  by four clinical research technicians at WFUSM (hereafter, ``auditors''), who were chosen for their familiarity with Epic (the institution's electronic charting software) and research experience.

\subsubsection{Identifying Sources of Auxiliary Information}

Before the chart reviews, we developed the ``audit roadmap'' (Table~\ref{tab:ali_components_roadmap}), harnessing clinical expertise to identify auxiliary information about missing ALI components. The roadmap suggested where auditors could look in Epic to locate supplemental information about missing values. It employs ``anchors'' to try to recover missing components using clinically-driven placeholders \cite{Halpern2014}.

For example, if a patient was missing HbA1c,  auditors would search their chart for a diabetes or impaired glycemic control diagnosis. If either was present, the patient's diagnosis would suggest that their missing HbA1c value would have exceeded the  $6.5\%$ threshold. Thus, their validated $X$ would include one point for HbA1c. If neither diagnosis was present, $X$ would exclude this still-missing component. 
 
Many components, like HbA1c, had multiple sources of auxiliary information, while others, like triglycerides, had just one. Only serum albumin, a blood test for liver and kidney function, did not have any. 

\subsubsection{Validated Data Collection}

In preparation, the extracted EHR data were transformed from a wide (one row per patient encounter) to a long format (one row per patient encounter per variable). This transformation streamlined the data structure for auditors, with one value per row to review (Supplemental Figure~S2).

\begin{figure}[ht]
\includegraphics[width=0.95\textwidth]{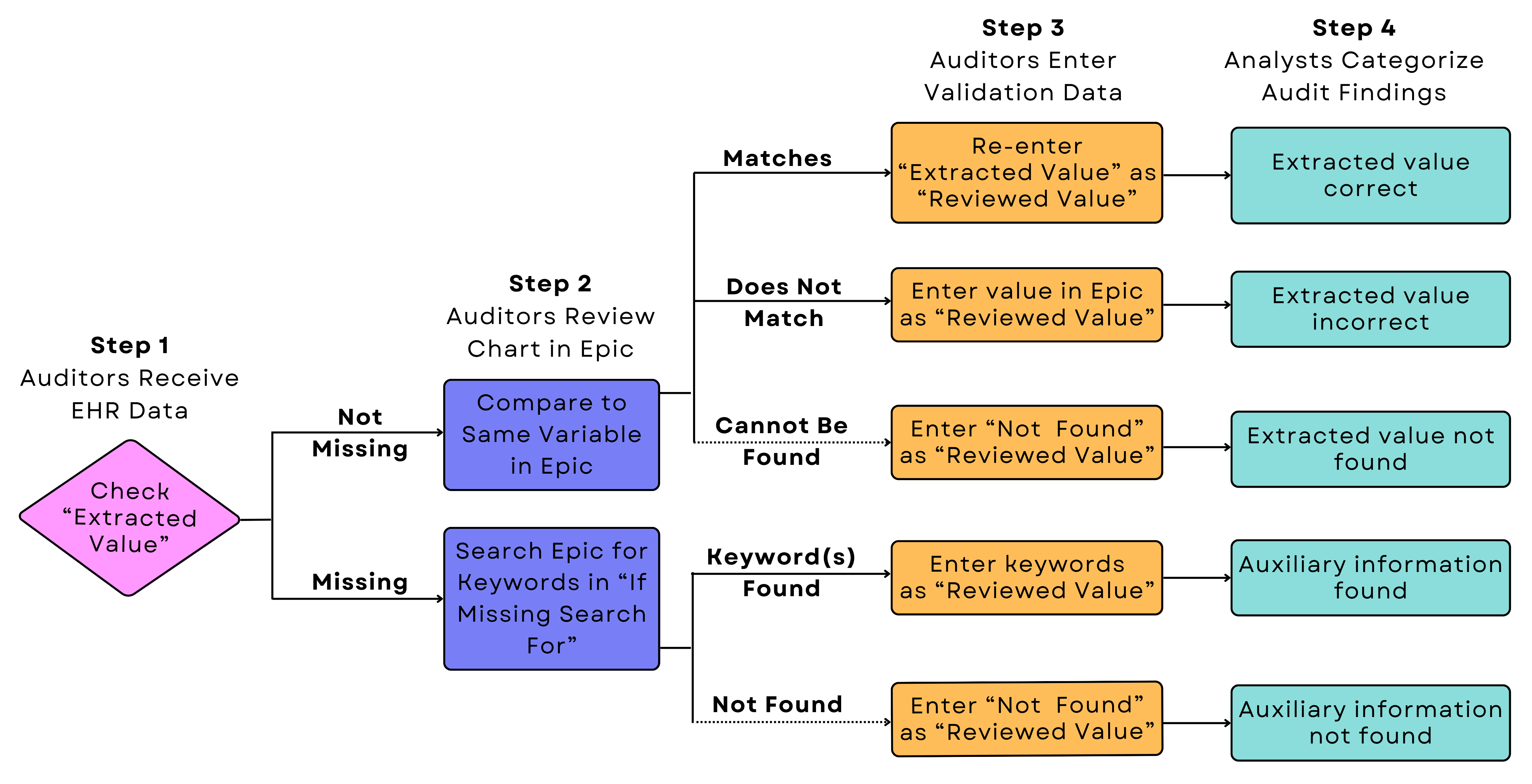}
\caption{Flow chart of instructions given to auditors for the chart reviews (Steps 1--3) and the corresponding auditor finding assigned by the analysts (Step 4). Column names refer to those in Supplemental Figure~S2.}\label{flow-instr}
\end{figure}

Before sending the data to the auditors, we incorporated the roadmap and added an empty column for the ``reviewed value'' (i.e., the one in Epic). Depending on the scenario, auditors would enter different information for the reviewed value, and each data point was assigned one of five auditor findings (Figure~\ref{flow-instr}). 

\subsection{Multi-Wave Strategy: Sampling Adaptively and Improving Tools} \label{subsec:multiwave}

Choosing the ``best'' design often relies on information that is unknown at baseline, like the EHR error rates or the logORs for the healthcare utilization model. Since we did not have any prior information about these unknowns, we started by sampling  conservatively to collect information for a targeted design later. This multi-wave validation scheme balances learning about our data with strategically designing for them. Even if prior information is available (e.g., from previous chart reviews), multi-wave validation allows us to learn how our data have changed before dedicating resources to validate them again \cite{lotspeichetal2024}. Further, we can incorporate auditors' feedback between waves to improve the validation protocol. 

We validated $100$ patients' data across three waves: the Pilot, Wave I, and Wave II. The Pilot wave tested the validation protocol and established inter-auditor agreement on the first four patients. Then, Waves I and II put this protocol into action as auditors validated the remaining patients. Waves I and II each selected $48$ more patients. (Designating equal-sized waves balances design flexibility and stability \cite{McIsaac&Cook2015}.) In Wave II, information from the Pilot and Wave I was leveraged to design a more targeted strategy. Patients could be chosen for validation at most once.

\subsubsection{Modifying the Validation Study Design as Information Accumulates}\label{subsec:adapt_design} 

Between the Pilot and Wave I, $52$ patients were selected for chart review via the BCC design. Specifically, patients were split into four strata based on their outcome and error-prone ALI (discretized at the median of $0.33$). Then, equal numbers of patients were randomly chosen from the four strata. In the Pilot, one patient was chosen from each stratum; in Wave I, another $12$ were. 

Before beginning Wave II, we designed simulation studies based on the preliminary validation data to explore how designs compared under different scenarios for true underlying data quality. The design offering the best efficiency (i.e., lowest variability) for the SMLE $\hat{\beta}_1$ of the logOR on ALI was desired. Details are in the Supplemental Materials.

Under varied error rates, the targeted designs offered empirically unbiased estimates and higher efficiency than SRS. The RS design provided the best efficiency under extra-low to moderate error settings, and the preliminary validation data (Pilot and Wave I) indicated a true positive rate (TPR) $= 1.00$ and a false positive rate (FPR) $< 0.01$.  

Under high missing data recovery rates ($90$--$100\%$) and different designs, the SMLE was empirically unbiased. While biased for moderate recovery or lower ($\leq 50\%$), the SMLE was still closer to the truth than the naive estimate was (i.e., based on unvalidated EHR data). Regardless of how much missing data were recovered, the RS design offered the highest efficiency.

We concluded that the RS design was the most informative way to select the remaining $48$ patients for validation in Wave II. Each unvalidated patient's residual was calculated based on the naive healthcare utilization model estimates (Section~\ref{results:mod}) following \eqref{def_resid} as 
\begin{align*}
r_i^* = Y_i - \frac{1}{1 + \exp\left\{-\left(-1.383 + 0.945X_i^* + 0.103Z_i\right)\right\}}.   
\end{align*} Then, patients were ranked by $r_i^*$. The $24$ patients with the smallest residuals and the $24$ patients with the largest residuals were selected for validation. Here, the RS design prioritized (i) patients with zero hospitalizations or ED visits but high predicted probabilities of having at least one and (ii) patients with at least one hospitalization or ED visit but high predicted probabilities of having zero. That is, patients whose $Y_i$ aligned with the naive model's prediction were not considered to be very informative and were deprioritized for validation.

\subsubsection{Testing the Validation Protocol and Establishing Inter-Auditor Agreement}\label{subsec:pilot} 

The validation protocol evolved slightly throughout the waves, as well. After completing the Pilot, auditors and analysts met to discuss any protocol modifications that would be helpful. For example, the auditor finding  ``extracted value not found'' was not originally an option, and we discussed how these data points should be recorded. Adding an optional auditor ``Notes'' column resulted from this conversation, too. 

We also evaluated inter-auditor agreement in the Pilot. We started with two auditors on our team. Each off them reviewed the four patients chosen for the Pilot, allowing us to calculate inter-auditor agreement in this doubly-validated sample. Later, two new auditors joined and also reviewed these four patients' records. Ultimately, agreement in the four auditors' findings was calculated using Fleiss' Kappa \cite{fleiss1971}, as detailed in the Supplemental Materials. We found high agreement for labs ($\kappa=0.90$) and vitals ($\kappa=0.81$). 

\section{Results}
\label{sec:app} 

\subsection{Cohort Description {and Study Objectives}} 

We considered adult patients ($18$--$65$ years old) with at least one primary care outpatient encounter at Atrium Health Wake Forest Baptist Hospital. The $1000$ sampled patients had a median age of $48$ years and a median income of \$52,320. They were $61\%$ female, $72\%$ White, and $6\%$ Hispanic. During the $2$-year study period, $318$ patients ($32\%$) had at least one hospitalization or ED visit. In the unvalidated EHR data, error-prone ALI was right-skewed (Supplemental Figure~S3). With a median of $0.33$ (interquartile range $= 0.17, 0.50$), patients experienced one out of every three non-missing components, on average. We validated ALI data for $100$ patients. 

First, we discuss the EHR data quality. Second, we present the healthcare utilization model. Within these sections, results are introduced chronologically, beginning with findings from the unvalidated EHR data and ending with those based on all validation waves. We also explore how these findings inform future implementations of the ALI in this healthcare system.

\subsection{Validation Study Findings: EHR Data Quality} 
\label{results:val} 

As expected, the extracted EHR data had lots of missingness (Supplemental Figure~S4). Rare inflammation labs were missing for $>95\%$ of patients (Supplemental Figure~S5A). Metabolic labs were missing for $21\%$ of patients. Cardiovascular components (from vitals) were non-missing for basically everyone. Fortunately, most patients ($78\%$) had at least five non-missing ALI components (Supplemental Figure~S5B). 

\subsubsection{Pilot and Wave I Chart Reviews}

\begin{figure}[ht]
    \centering
    \includegraphics[width=\linewidth]{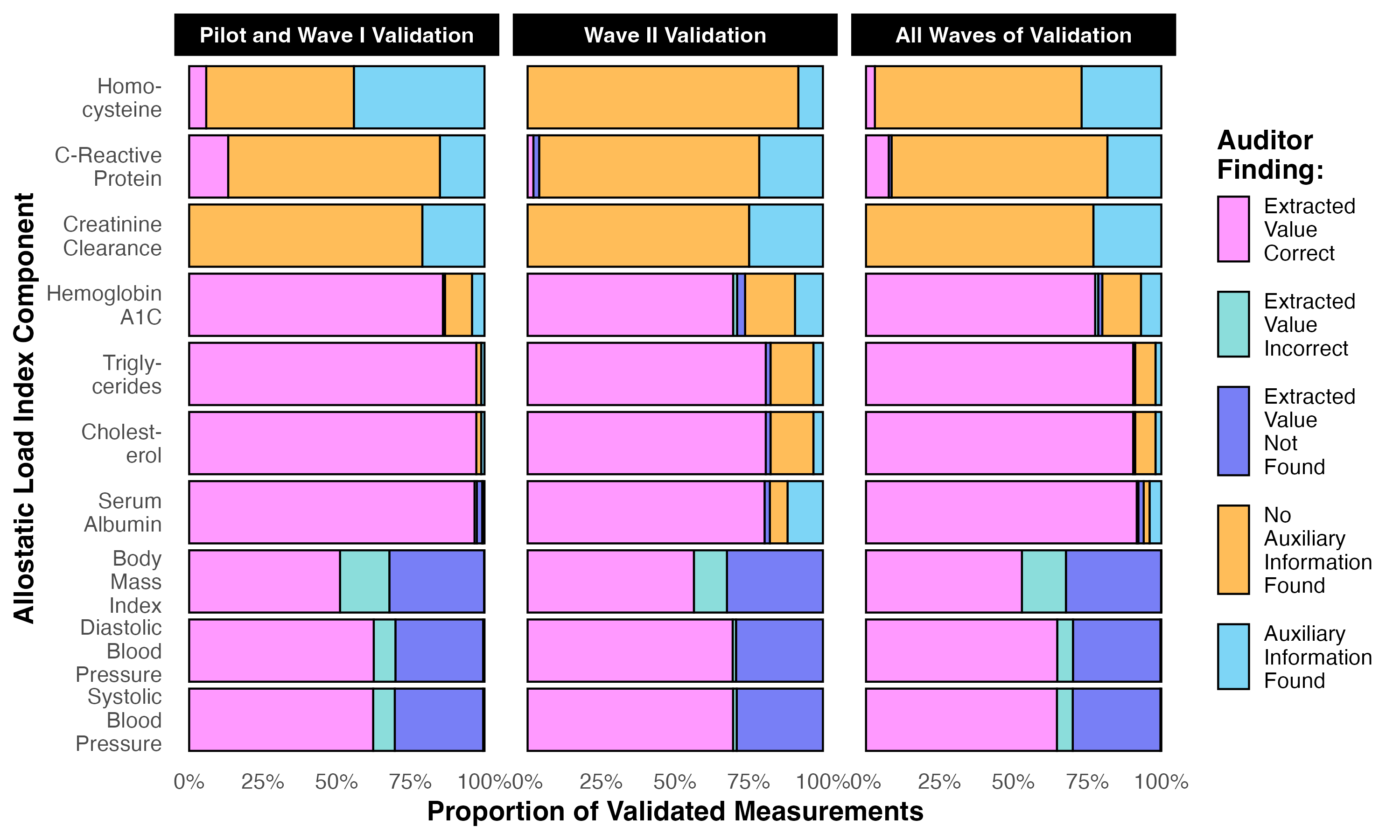}
    \caption{Auditors' findings from the validation of \textbf{A)} $52$ patients' data in the Pilot and Wave I, \textbf{B)} $48$ patients' data in  Wave II, and \textbf{C)} $100$ patients' data in all waves combined. These findings refer to the original numeric measurements (before discretizing them into allostatic load index components), and there could be multiple per patient per variable. Auditor findings were mutually exclusive and specific to whether the measurements were originally \textit{non-missing} (``extracted value correct,'' ``extracted value incorrect,'' or ``extracted value not found'') or \textit{missing} (``no auxiliary information found'' or ``auxiliary information found'') in the extracted electronic health records data.} 
    \label{fig:findings}
\end{figure}

\begin{figure}[ht]
    \centering
    \includegraphics[width=\linewidth]{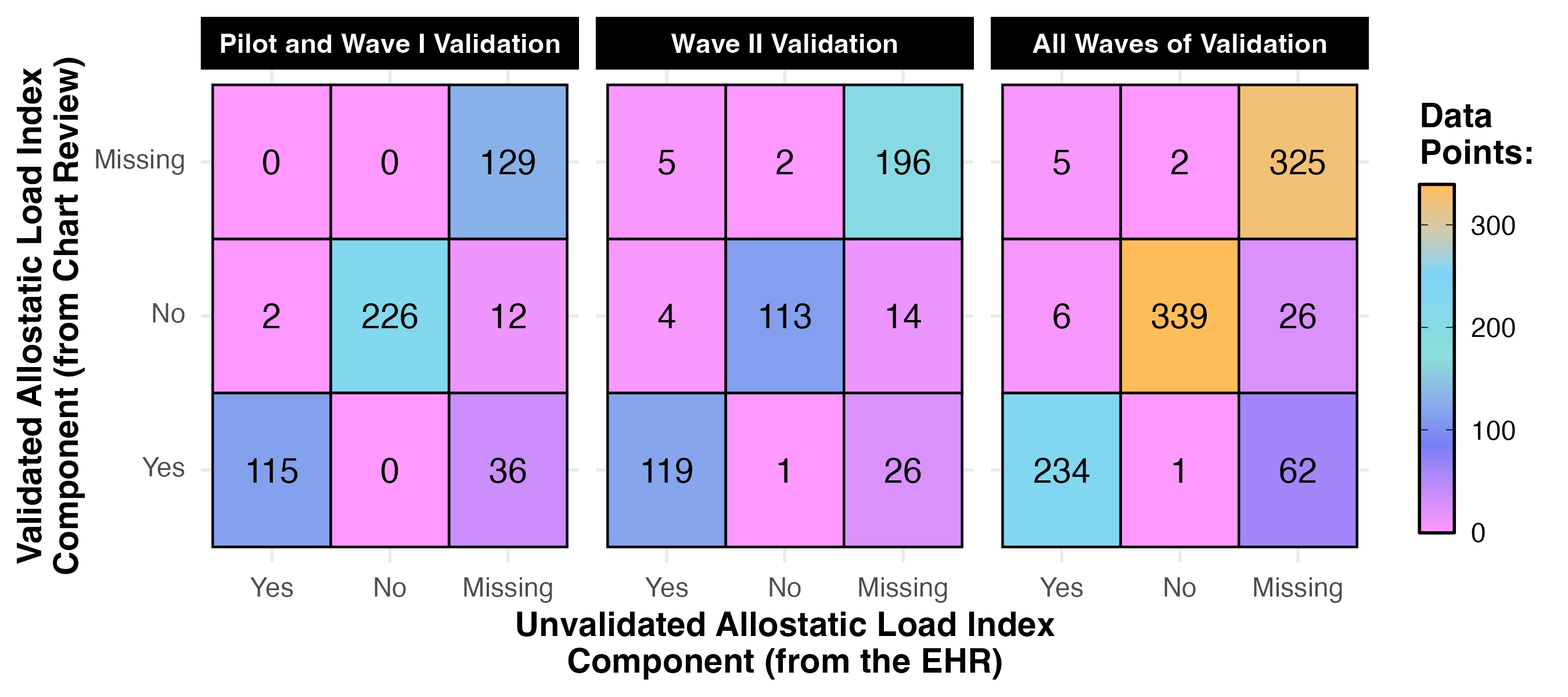}
    \caption{Comparison of the discretized allostatic load index components from the unvalidated electronic health records (EHR) data to \textbf{A)} $52$ patients' data in the Pilot and Wave I, \textbf{B)} $48$ patients' data in  Wave II, and \textbf{C)} $100$ patients' data in all waves combined.}
    \label{fig:heatmap}
\end{figure}

In total, $\num{7605}$ data points were validated ($\num{1049}$ labs, $\num{6556}$ vitals) between the Pilot and Wave I (Figure~\ref{fig:findings}A). Patients had $13$--$763$ data points validated each (median $= 81.5$). For common labs, most extracted values matched Epic, and all non-missing values for rare labs matched (Supplemental Table~S1). Only two non-missing ALI components were misclassified (Figure~\ref{fig:heatmap}A), for an estimated TPR $=100\%$ and FPR $<1\%$. 

Fewer vitals than labs matched Epic ($59\%$ versus $81\%$), with $31\%$ not found. We discovered that the EHR data extraction algorithm expected every patient encounter to have vitals. When vitals were not taken (e.g., physical therapy), values were imputed from other encounters; auditors could not find these values in Epic. Thus, we uncovered an improvement area in the EHR data extraction protocol. 

Incorporating the roadmap into the validation protocol improved the completeness of the data. The number of missing values decreased for nearly all ALI components (Supplemental Figure~S5A). Homocysteine saw the biggest improvement, with auxiliary information located for $23$ of  $52$ validated patients ($44\%$). The median number of non-missing ALI components per patient increased from $7$ to $8$. Overall, $48$ of $177$ missing components ($27\%$) were recovered (Figure~\ref{fig:heatmap}A).

\subsubsection{Wave II Chart Reviews}

In Wave II, $3867$ data points were validated ($718$ labs, $3149$ vitals), which was fewer than in the Pilot and Wave I. This difference was explained by fewer patient encounters, with $16$--$453$ validated data points per patient (median $= 65.5$). The estimated error rates for non-missing components (TPR $=99\%$, FPR $=3\%$) and missing data recovery rate ($17\%$) in Wave II closely resembled the Pilot and Wave I (Figure~\ref{fig:heatmap}B). Still, some auditors' findings differed  (Figure~\ref{fig:findings}B), like more vitals matching Epic ($65\%$). 

Combined findings from all validation waves indicated that non-missing EHR data were of excellent quality (TPR $= 99.5\%$, FPR $= 2\%$), and auxiliary information was located for $21\%$ of missing components (Figures~\ref{fig:findings}C and \ref{fig:heatmap}C). The median number of non-missing components per patient increased from $6$ to $7$. Supplemental Tables~S1--S2 contain detailed findings. 

\subsection{Analysis Findings: Whole-Person Health and Healthcare Utilization} 
\label{results:mod} 

Using the unvalidated EHR data, we first fit the naive model. The expected odds of engaging in the healthcare system for an $18$-year-old patient with an ALI of $0$ were $0.25$ (95\% confidence interval [95\% CI]: $0.18$, $0.36$). For every $0.1$-point (one-component) increase in ALI, these odds were expected to increase by $10\%$ (OR $=1.10$, 95\% CI: $1.03$, $1.18$), adjusting for age. For every $10$ years older a patient was, their odds were expected to increase by $11\%$ (OR $=1.11$, 95\% CI: $1.00, 1.23$), adjusting for ALI. While these findings align with the literature, we wanted to ensure that missingness and potential errors in the EHR data did not lead to spurious conclusions. 

\subsubsection{Preliminary Estimates Based on the Pilot and Wave I Validation}
 
Incorporating the Pilot and Wave I, we re-fit the model with the first $52$ patients' validated data and all $1000$ patients' unvalidated data. Simulations in Section~\ref{subsec:adapt_design} were designed from these ``preliminary SMLEs.'' For an 18-year-old patient with an ALI of $0$, the expected odds of engaging in the healthcare system decreased to $0.17$ (95\% CI: $0.12$, $0.26$). The association between ALI and healthcare utilization appeared even stronger, with an estimated 21\% increase in odds for each $0.1$-point increase (OR $=1.21$, 95\% CI: $1.17, 1.25$), adjusting for age. The adjusted OR $= 1.10$ for age was relatively unchanged (OR $= 1.10$, 95\% CI: $1.01, 1.21$). Preliminary SMLEs reaffirmed that worse whole-person health was associated with higher expected odds of engaging in the healthcare system, perhaps more strongly than the naive model indicated.

\subsubsection{Final Estimates Based on All Waves of Validation}

After completing all three waves, we used the $100$ patients' validated data with all $1000$ patients' unvalidated data to re-fit the SMLEs one last time. These final estimates all fell between the naive model and preliminary SMLEs. The expected odds of engaging in the healthcare system for an 18-year-old patient with an ALI of $0$ were $0.23$ (95\% CI: $0.16$, $0.33$). For each $0.1$-point increase in ALI, a $12\%$ increase in odds was expected (OR $= 1.12$, 95\% CI: $1.09, 1.15$), adjusting for age. Again, older patients were expected to be more likely to engage (adjusted OR $= 1.11$, 95\% CI: $1.02, 1.20$). Technical details and R code are in the Supplemental Materials.

\subsection{Informing Implementation in the Learning Health System}\label{res:inform} 

All models agreed that worse whole-person health (higher ALI) was associated with a higher likelihood of engaging in the healthcare system, adjusting for age. Chart reviews uncovered high quality of non-missing EHR data and located auxiliary information for some missing data. Without validation, knowing whether we could trust the EHR data and whether the EHR-derived ALI was a valid computable phenotype for whole-person health would not have been possible. Thus, while the models led to the same conclusions, validating EHR data before using them in practice or research is critical. 

Our findings can now inform implementation strategies to incorporate the ALI into this learning health system. For example, if the ALI automatically prepopulates in patients' charts, clinicians would have easy access to it when making care decisions, and researchers could extract it for further studies. This healthcare system previously implemented another score, the EFI, across the patient population aged $65$ and older. Building on that experience and encouraged by the ALI's predictive ability, we plan to operationalize it for patients aged $18$ and older.

Of course, we want to embed the validated ALI, and only a small subset of patients can undergo chart reviews. Fortunately, in obtaining SMLEs for the healthcare utilization model, we robustly estimate the exposure error mechanism, $\Pr(X|X^*,Z)$. When scaling up across the healthcare system, the estimated exposure error mechanism can be used to efficiently predict patients' validated ALI from unvalidated EHR data (Figure~\ref{fig:pred_ali}). See the Supplemental Materials for details on calculating these predictions.

\begin{figure}[ht]
    \centering
    \includegraphics[width=\linewidth]{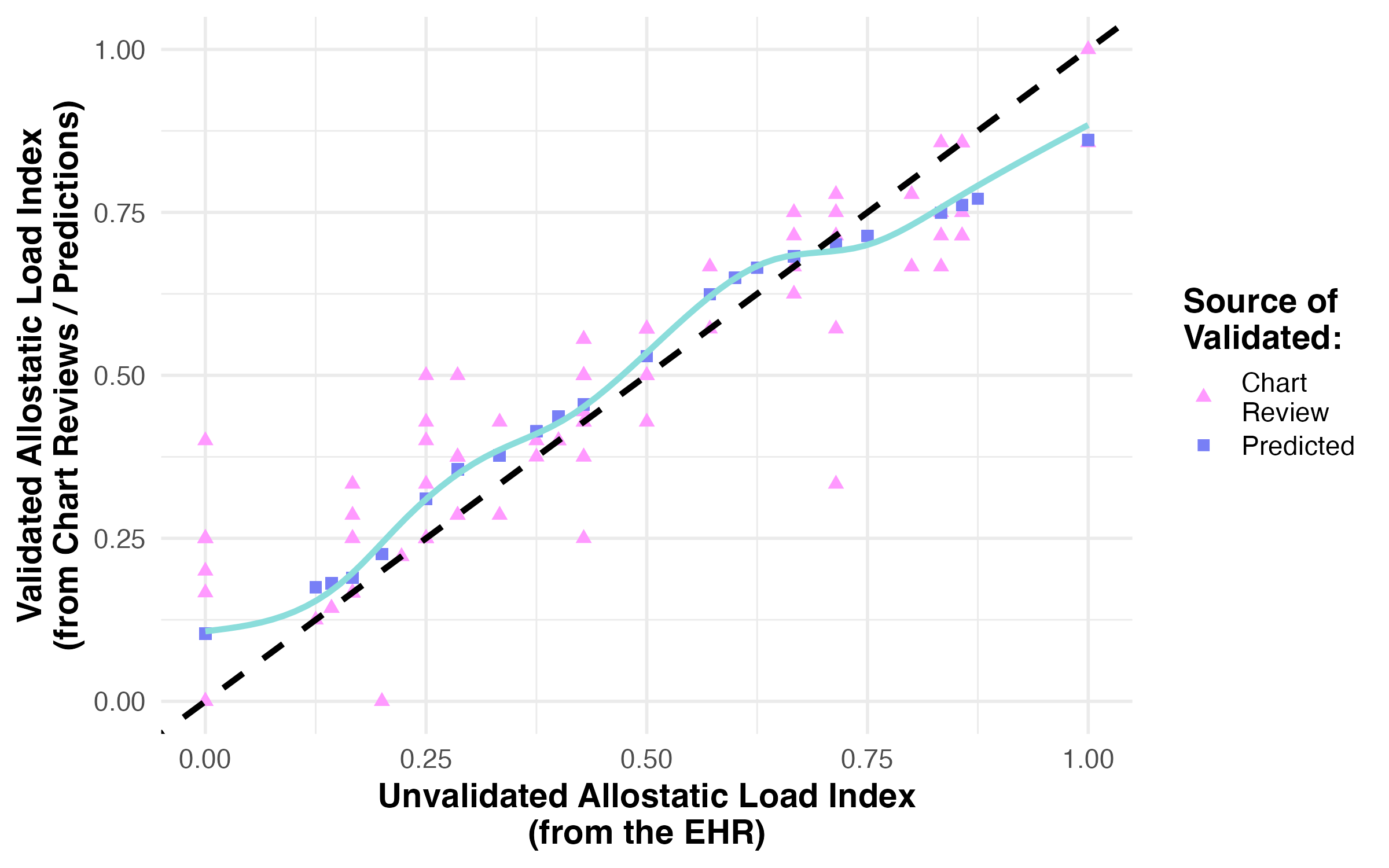}
    \caption{Among patients with chart reviews (triangles), validated allostatic load index (ALI) could differ noticeably from the unvalidated version using the unvalidated electronic health records (EHR) data. For patients without chart reviews (squares), validated ALI could be robustly predicted using the estimated exposure error mechanism. The dashed line indicates the line of equality (i.e., where unvalidated and validated ALI are the same). The solid line is a loess smoother.}
    \label{fig:pred_ali}
\end{figure}

\section{Discussion}
\label{sec:disc}

Predicting healthcare utilization is a priority in the learning health system, with important implications at both patient and institutional levels. Establishing standardized, valid measures of whole-person health (like the ALI) that are associated with utilization and can scale up to the entire institution is an integral step. While EHR data contain variables to calculate these measures, missing data is a key challenge, and non-missing data are still prone to errors. 

Data validation, often through chart reviews, can gauge data quality, but this time-intensive process is usually only possible for a small subset of patients. Also, existing validation protocols do not overcome missingness. We propose novel methods to maximize information gained from a partial validation study through the statistical model, validation protocol, and targeted patient selection strategy. We focused on estimating the association between ALI and healthcare utilization. In preparation for EHR implementation, partial validation can also efficiently improve estimates of the ALI itself. 

The proposed analytical and design strategies were devised to ensure robust and efficient estimates. Specifically, the statistical model avoids making restrictive assumptions about the error mechanism,  while incorporating all available information to achieve precise estimates (i.e., low variance). Moreover, the targeted design strategies are tailored to the preliminary validation data and chosen statistical model, selecting patients who are most informative for the clinical question. Our enriched validation protocol builds on the literature (e.g., \cite{Duda2012}) by incorporating a clinically-driven ``roadmap'' to locate auxiliary information about missing values. 

Due to the curse of dimensionality, the SMLEs can handle conditioning on error-prone $X^*$ and maybe one other covariate $Z$ in the error mechanism. We assumed that validated ALI depended only on its unvalidated version. However, if a more complicated error structure is expected, data pre-processing will be needed before using the SMLEs \cite{Tao2021}.

While our roadmap recovered some missing ALI components, {validated patients could still have incomplete data}. Clinical guidelines were only defined for missing values presumed to be unhealthy. Missingness could be further reduced if auditors were given additional guidance on auxiliary information for healthy values. Others have described missingness in the EHR as selection bias and corrected for it using approaches like IPW \cite{Peskoe2021}. Applying such approaches to our post-validation data may reduce the impact of residual missingness. Alternatively, we could adopt other allostatic load calculations, like the index of cardiometabolic health, which rely less on rare labs and may face fewer missing components \cite{Nobel2017}. 

EHR data are susceptible to ``informative presence bias,'' wherein the process through which patients engage in the healthcare system can bias associations \cite{Goldstein2019, Harton2022, McGee2022}. Our decision to adapt the Seeman et al. (2001) \cite{Seeman2001} ALI as the proportion of non-missing components could be considered a summary measure. A recent review found summary measures to be the most common approach to handling informative presence bias \cite{Sisk2021}.

The ALI is among many computable phenotypes currently applied to EHR data \cite{Ting2023}. Other phenotyping algorithms, including EHRshot \cite{Wornow2023}, translate observable  characteristics into measures of underlying patient states. Validation is a critical step in their development and typically involves chart reviews to obtain ``gold standard'' labels \cite{GRAHAM2022100974, McDonough2020}. When phenotyping with machine learning methods, this validation set generally needs to be an SRS. Meanwhile, our statistical model accommodates targeted validation sampling, creating opportunities to promote the efficiency of downstream analyses (e.g., associations between the phenotype and suspected risk factors). 

There are several interesting directions for future work. First, a longitudinal approach to modeling healthcare utilization, wherein patients' ALI can vary over time, might provide more information. Second, auditors recorded their findings in Microsoft Excel, but an electronic REDCap \cite{Harris2009} interface would improve usability and accuracy  \cite{Lotspeich2023}. Third, targeted validation strategies could be used to improve predictive ability rather than efficiency \cite{Liu2022, TanHeagerty2015}. 

\section{Conclusion}
\label{sec:concl} 
EHR data provide an exciting opportunity to implement key measures in the learning health system. However, errors and missingness in these data pose challenges in verifying and operationalizing such measures. Validation of even a subset of patients' data is time- and cost-intensive but provides crucial insights into data quality. Statistical and biomedical informatics methods can maximize information gained from such partial validation. 

\bibliographystyle{wileyNJD-AMA}
\bibliography{Bibliography-MM-MC}





\section*{Acknowledgments}
The authors gratefully acknowledge Wake Forest University and Wake Forest University School of Medicine for an intercampus collaborative grant that supported this work. Computations were performed using the Wake Forest University (WFU) High Performance Computing Facility, a centrally managed computational resource available to WFU researchers including faculty, staff, students, and collaborators. 

\section*{Supplementary Materials}
\begin{itemize}
    \item \textbf{Additional appendices, tables, and figures:} The supplemental figures and tables referenced in Sections 2--3 are available online at \url{https://github.com/sarahlotspeich/ALI_EHR/blob/main/Supplement.pdf} as Supplementary Materials.
    \item \textbf{R-package for the SMLE:} An \textsf{R} package \texttt{logiSieve} that implements the semiparametric sieve maximum likelihood estimators described in this article is available at \url{https://github.com/sarahlotspeich/logiSieve}.
    \item \textbf{R-package for sampling designs:} An \textsf{R} package \texttt{auditDesignR} that implements the various validation study designs in this article is available at \url{https://github.com/sarahlotspeich/auditDesignR}.
    \item \textbf{R code and data for simulation studies:} The R scripts and data needed to replicate the simulation studies are available at \url{https://github.com/sarahlotspeich/ALI_EHR}.
\end{itemize}


\end{document}